\documentclass[12pt,aps]{revtex4}
\pdfoutput=1

\usepackage{graphicx}
\begin{document}

\title{Multi-channel modelling of the formation of vibrationally cold polar KRb molecules.} 

\author{Svetlana Kotochigova$^1$, Eite Tiesinga$^2$, and Paul S. Julienne$^2$}

\vspace*{0.5cm}

\affiliation{$^{1}$ Department of Physics, Temple University, Philadelphia, PA 19122-6082, USA \\ 
$^{2}$ Joint Quantum Institute, NIST and University of Maryland, Gaithersburg, Maryland 20899-8423, USA.}

\begin{abstract}
We describe the theoretical advances that influenced the
experimental creation of  vibrationally and translationally cold polar
$^{40}$K$^{87}$Rb molecules  \cite{nphys08,science08}. Cold molecules
were created from very-weakly bound molecules formed by magnetic
field sweeps near a Feshbach resonance in collisions of ultra-cold
$^{40}$K and $^{87}$Rb atoms. Our analysis include the multi-channel
bound-state calculations of the hyperfine and Zeeman mixed X$^1\Sigma^+$
and a$^3\Sigma^+$ vibrational levels.  We find excellent agreement with
the hyperfine structure observed in experimental data.  In addition,
we studied the spin-orbit mixing in the intermediate state of the
Raman transition. This allowed us to investigate its effect on
the vibrationally-averaged transition dipole moment to  the lowest
ro-vibrational level of the X$^1\Sigma^+$ state. Finally, we obtained an
estimate of the polarizability of the initial and final ro-vibrational
states of the Raman transition near frequencies relevant for optical
trapping of the molecules.
\end{abstract}

\maketitle

\section{Introduction}

The recent successful creation of a high phase-space-density gas of polar
$^{40}$K $^{87}$Rb molecules \cite{nphys08,science08} has been based
on both new experimental and theoretical advances in manipulating and
understanding properties of such molecules.  This opens up the possibility
of studying collective phenomena that rely on the long-range interactions
between polar molecules.  Future experiments can be envisioned in both
weakly confining optical traps as well as optical lattices.

Our goal  in this paper is to describe some of the theoretical
advances that influenced the experimental creation of  vibrationally and
translationally cold polar $^{40}$K$^{87}$Rb molecules.   In particular,
we theoretically analyze various factors that can affect this creation
including the multi-channel description of the initial, intermediate,
and final states of the formation by Raman transitions.

In a previous paper \cite{Kotoch03} we made the first steps towards
obtaining practical guidelines for photoassociatively producing low $v$
vibrational states of heteronuclear KRb.  We calculated the electronic
transition dipole moments between ground state of the KRb molecule
and excited states.  In addition, we obtained the permanent dipole
moments of the polar X$^{1}\Sigma^+$ and $a^{3}\Sigma^+$ ground states.
A relativistic electronic structure code was used.

In a second paper \cite{Kotoch04} we discussed the possibility of creating
X$^{1}\Sigma^+$ molecules starting from doubly spin-polarized K and Rb
atoms via two-photon photoassociation.  We assumed that colliding atoms
are initially in the doubly spin-polarized state, which only allows
them to bond in the ground configuration $a^{3}\Sigma^+$ potential.
We then found that there is a viable route from the doubly spin-polarized
colliding atoms to the vibrationally ground $X^{1}\Sigma$ state via the
excited A$^1\Sigma^+$ and b$^3\Pi$ states, when they are mixed through
spin-orbit interactions. This mixing allows us to describe these potential
by the Hund's case (c) coupling scheme as 2($0^+$) and 3($0^+$) states
and calculate the Raman transition rate from the triplet state to the
excited $\Omega = 0^+$ state followed by a downward transition to the
ground singlet state.  This process is absent in homonuclear dimers,
since the additional {\it gerade-ungerade} symmetry prevents it. In
the notation m($\Omega$) the number in parenthesis is the projection
$\Omega$ of the total electronic angular momentum on the internuclear
axis and the number in front labels the order of states with the same
$\Omega$. The role of black-body radiation in redistributing population
among ro-vibrational levels of the singlet X$^1\Sigma^+$ and triplet
a$^3\Sigma^+$ states was investigated.

Here we further search for an efficient production mechanism using a
multi-channel description of both ground and excited states.  We assume
that KRb molecules are initially in the weakly-bound near-threshold
vibrational states formed by a magnetic Feshbach resonance in collisions
between ultracold $^{40}$K and $^{87}$Rb atoms. In our coupled-channel
calculation of the ground state ro-vibrational structure we used the most
accurate ground state potentials available from Ref.~\cite{Tiemann}.
In section~\ref{a} we analyze this structure, and perform a comparison
with available high-precision measurements \cite{nphys08,science08}.
\begin{figure}
\includegraphics[scale=0.35]{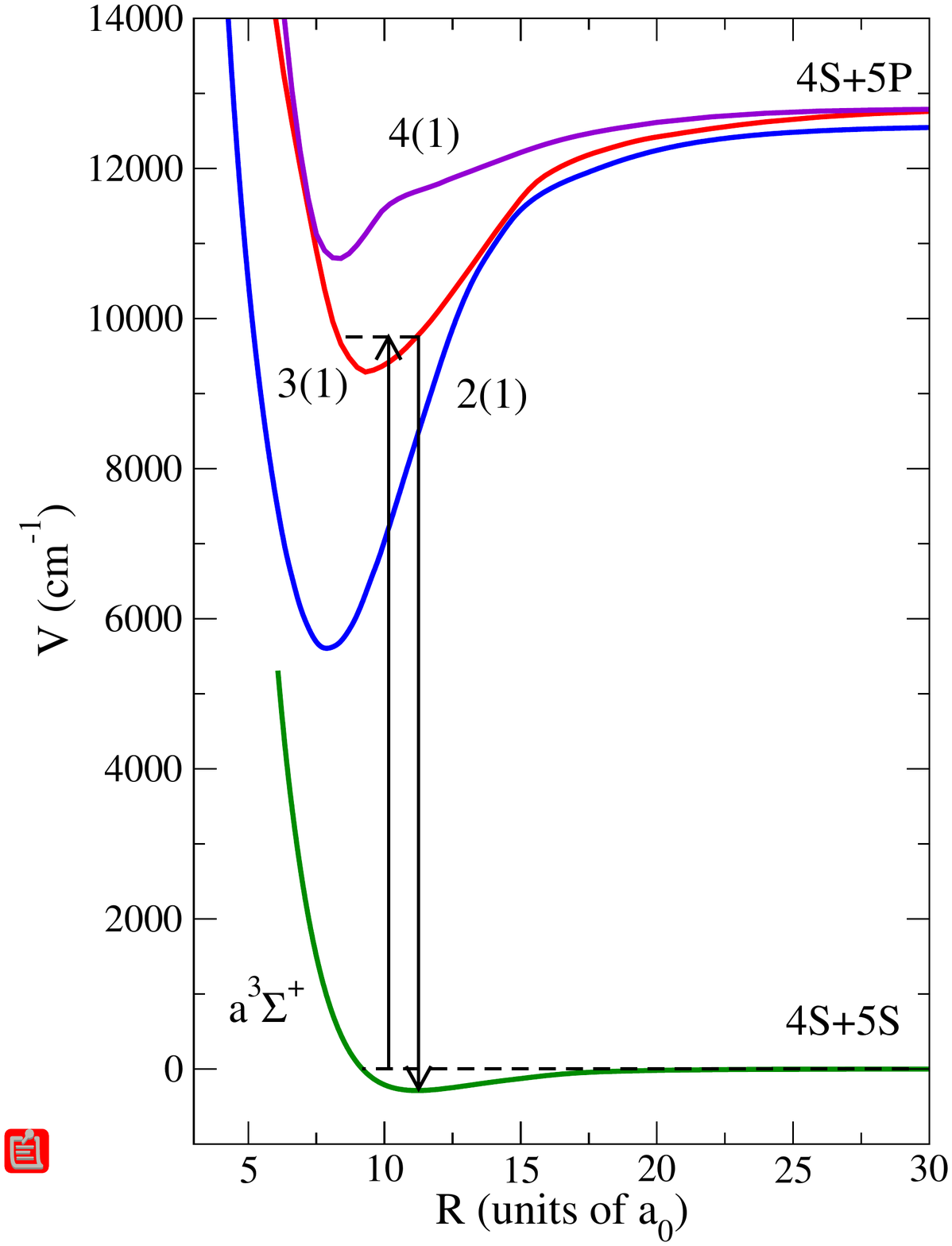}
\includegraphics[scale=0.35]{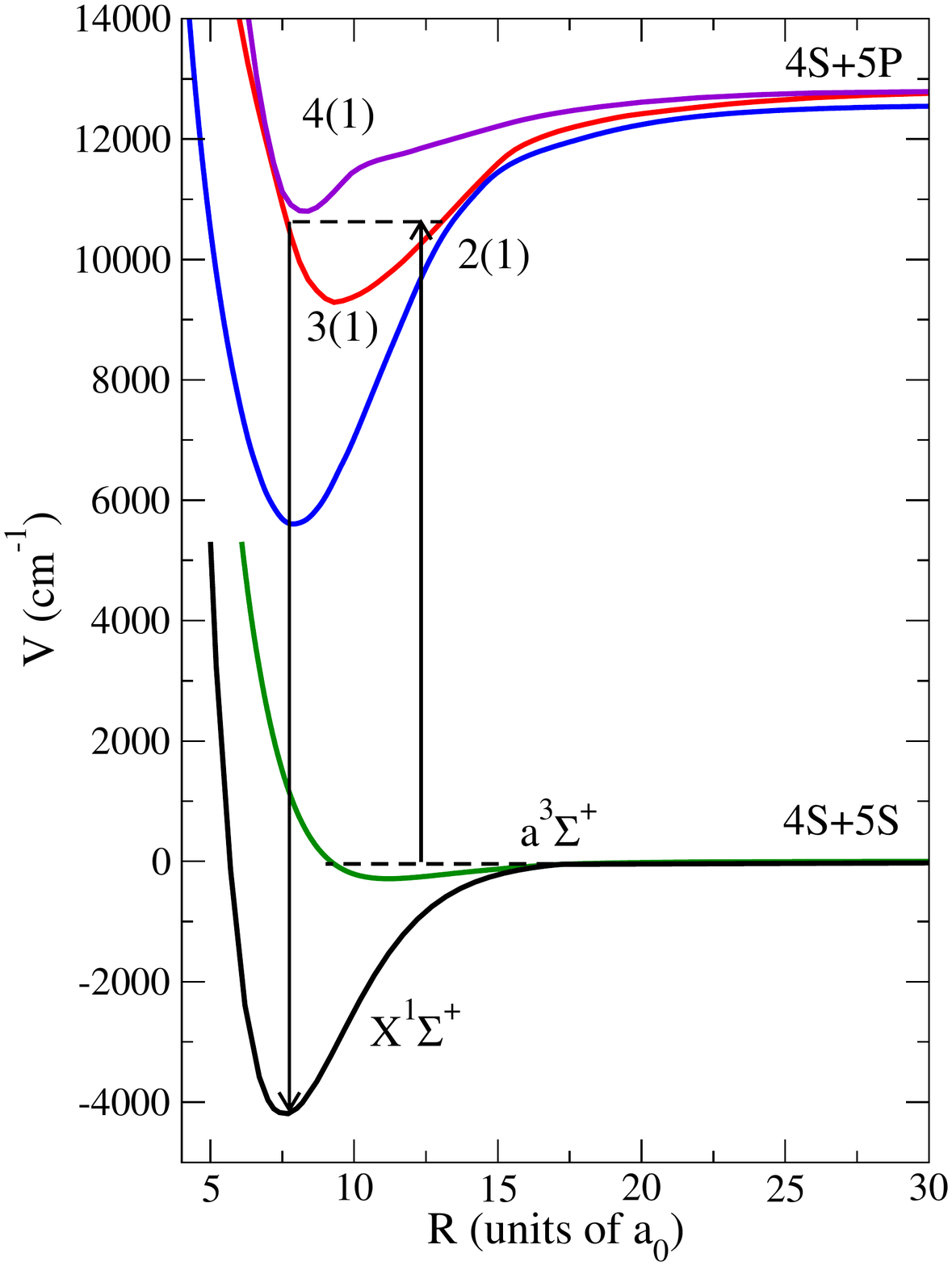}
\caption{The ground and excited state potential energy curves of KRb 
that are used for multi-channel modeling as a function of internuclear separation $R$. Here $a_0$
is the Bohr radius of 0.0529 nm.}
\label{scheme}
\end{figure}

Vibrationally cold molecules are preferably made by transfering population
from a Feshbach molecular state. For this transfer we selected the
pathway that has been proposed by Stwalley \cite{Stwalley}, which
forms vibrationally cold KRb molecules starting from the highly excited
vibrational states using one optical Raman transition and intermediate
vibrational levels of the 3(1) potential.  This mechanism was also used
to create vibrationally cold RbCs molecules in Ref.~\cite{DeMille}.
Reference~\cite{Bergeman} has reported an analysis of perturbations of the
vibrational levels of the 3(1) potential due to spin-orbit interactions
with the neighboring potentials in RbCs.

For the vibrational levels used as intermediate states the 3(1) potential
can to first order be described as the nonrelativistic  2$^3\Sigma^+$
state. More accurately  the strong non-adiabatic interaction with the
neighboring 2(1) and 4(1) potentials has to be taken into account.
Alternately, we can view this coupling as being due to the spin-orbit
interaction between the nonrelativistic 2$^3\Sigma^+$, 1$^3\Pi$,
and 1$^1\Pi$ potentials.  Therefore in Section~\ref{b} we perform
multi-channel calculations of the ro-vibrational structure and the
vibrationally-averaged transition dipole moments to the ground state
levels. The three intermediate excited states that are of interest have
$\Omega$ = 1 symmetry and are shown in Fig.~\ref{scheme} together with the
ground state potentials of KRb. In addition, panel a of Fig.~\ref{scheme}
shows the pathway to form $v=0$ a$^3\Sigma^+$ molecules starting from a
gas of Feshbach molecules. Similarly, panel b shows the pathway to $v=0$
X$^1\Sigma^+$ molecules.

The excited potentials in Fig.~\ref{scheme} were  constructed from RKR
data \cite{Kasahara,Amiot1} and as well as from our {\it ab~initio}
calculations.  The two attractive 3(1) and 4(1) potentials dissociate
to the K(4$s$)+Rb(5$p_{3/2}$) atomic limit whereas the 2(1) potential
dissociates to the K(4$s$)+Rb(5$p_{1/2}$) limit.  At short internuclear
separations the potentials can be approximately described by the
Hund's case (a) $n^{2S+1}\Lambda^{\pm}$ symmetry, where $\Lambda$
is the projection of the electron orbital angular momentum along the
internuclear axis and $S$ is the total electron spin. At longer $R$
relativistic effects are important, where the curves can only be described
with Hund's case (c) $m(\Omega=1)$ labeling.

In Section~\ref{b} our multi-channel calculation is, however, based on the
nonrelativistic excited potentials 1$^3\Pi$, 1$^1\Pi$, and 2$^3\Sigma$,
electronic transition dipole moments to the ground states, and the
spin-orbit coupling matrix elements. Some RKR data for the more deeply
bound vibrational levels of the 1$^1\Pi$ and 2$^3\Sigma$ are available
\cite{Kasahara,Amiot1}.  We extend this information by {\it ab~initio}
electronic structure data and by known long range dispersion coefficients
from Ref.~\cite{Bussery}.

The relativistic configuration interaction molecular orbital restricted
active space (MOL-RAS-CI) method has been used to calculate potential
energy curves, permanent and transition electric dipole moments of
the KRb heteronuclear molecule as a function of internuclear separation.
We combine this calculation with multi-channel ro-vibrational structure
calculation to obtain Frank-Condon factors between the ground and excited
states of KRb.

Finally, in Section~\ref{c} we describe our calculation of the
polarizability of vibrational levels of the X$^1\Sigma^+$ and a$^3\Sigma^+$
states.

\section{Coupled-channel calculation of the ground states}\label{a}

In the Raman transition the initial and final bound vibrational levels
belong to the ground  X$^1\Sigma^+$ and a$^3\Sigma^+$ states. In
KRb these states dissociate to the same [Ar]4s($^2$S)+[Kr]5s($^2$S)
atomic limit. The two states are coupled via hyperfine interactions: the
Fermi-contact and electron and nuclear Zeeman interactions for each of
the constituent atoms. For each atom the Fermi-contact interaction couples
its electron spin, here 1/2, to its nuclear spin. The Zeeman interaction
is non-zero since an external magnetic field is used to create Feshbach KRb
molecules. The corresponding Hamiltonian for vibrational states in this
coupled system has been discussed in Refs.~\cite{Stoof,Tiesinga}. For
this paper we include the effect of the weak magnetic dipole-dipole
interaction perturbatively but do not include the effect of the
second-order spin-orbit interaction \cite{Mies}.

Numerically solving for eigen pairs of this system is called a
coupled-channel calculation. For our calculations we used the electronic
potentials of Ref.~\cite{Tiemann}. The atomic masses for $^{40}$K and
$^{87}$Rb are taken from  Ref.~\cite{Audi}, the Fermi-contact term values
and electronic g-factors are from Ref.~\cite{Arimondo} and, finally,
the nuclear magnetic moments are from Ref.~\cite{Stone}.  The nuclear
spin of $^{40}$K is 4 and that of $^{87}$Rb is 3/2.

The vibrational wave functions can be labeled by three nearly conserved
angular momentum quantum numbers. These are the relative orbital angular
momentum $\ell$ of the two atoms, its projection $m_\ell$ along the
magnetic field direction, and the projection $M_F$ of the summed
atomic angular momentum $\vec F=\vec f_a +\vec f_b$ along the same
direction. Here, $\vec f_a$ and $\vec f_b$ are total angular momenta of
each atom.  In fact, we can also write $\vec F =\vec S +\vec I$,
where $\vec S$ and $\vec I$ are the total electron and nuclear spin, respectively.

For our calculations we neglect coupling between
states with different $\ell\,m_\ell\,M_F$.  In the absence of the magnetic
dipole interaction the $2\ell+1$ levels with the same $\ell$ and $M_F$
quantum numbers are degenerate. For ultracold atoms and molecules we can
limit ourselves to $\ell=0$ and 2.  As the Feshbach molecule created
in \cite{Zirbel,science08} has $\ell=0$ and $M_F=-7/2$  we will limit
ourselves to $M_F=-11/2$ to $-3/2$ as in a Raman transition $M_F$ can 
change by upto two units depending on the polarization of the light beams.

\begin{figure}
\includegraphics[scale=0.35]{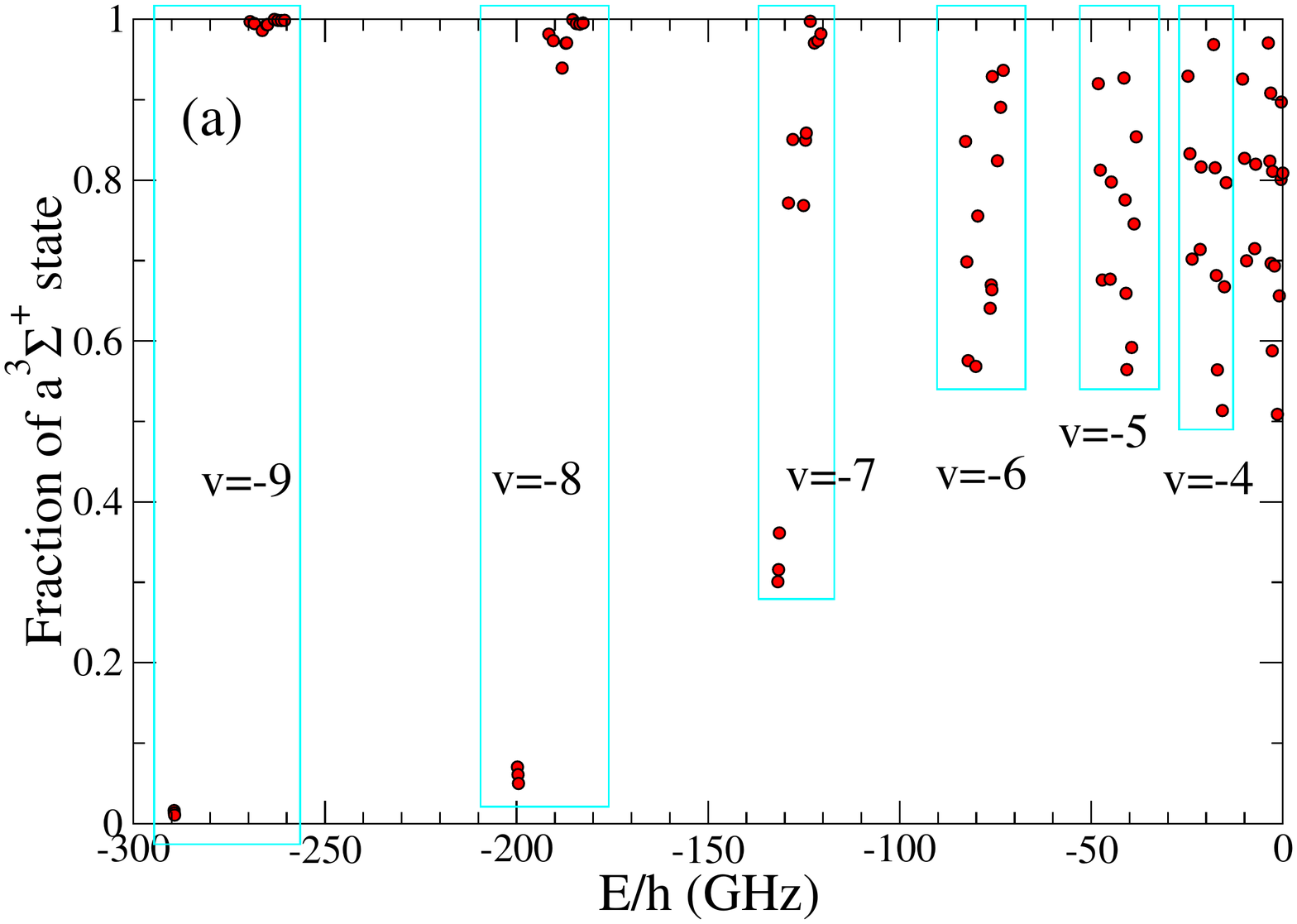}
\includegraphics[scale=0.35]{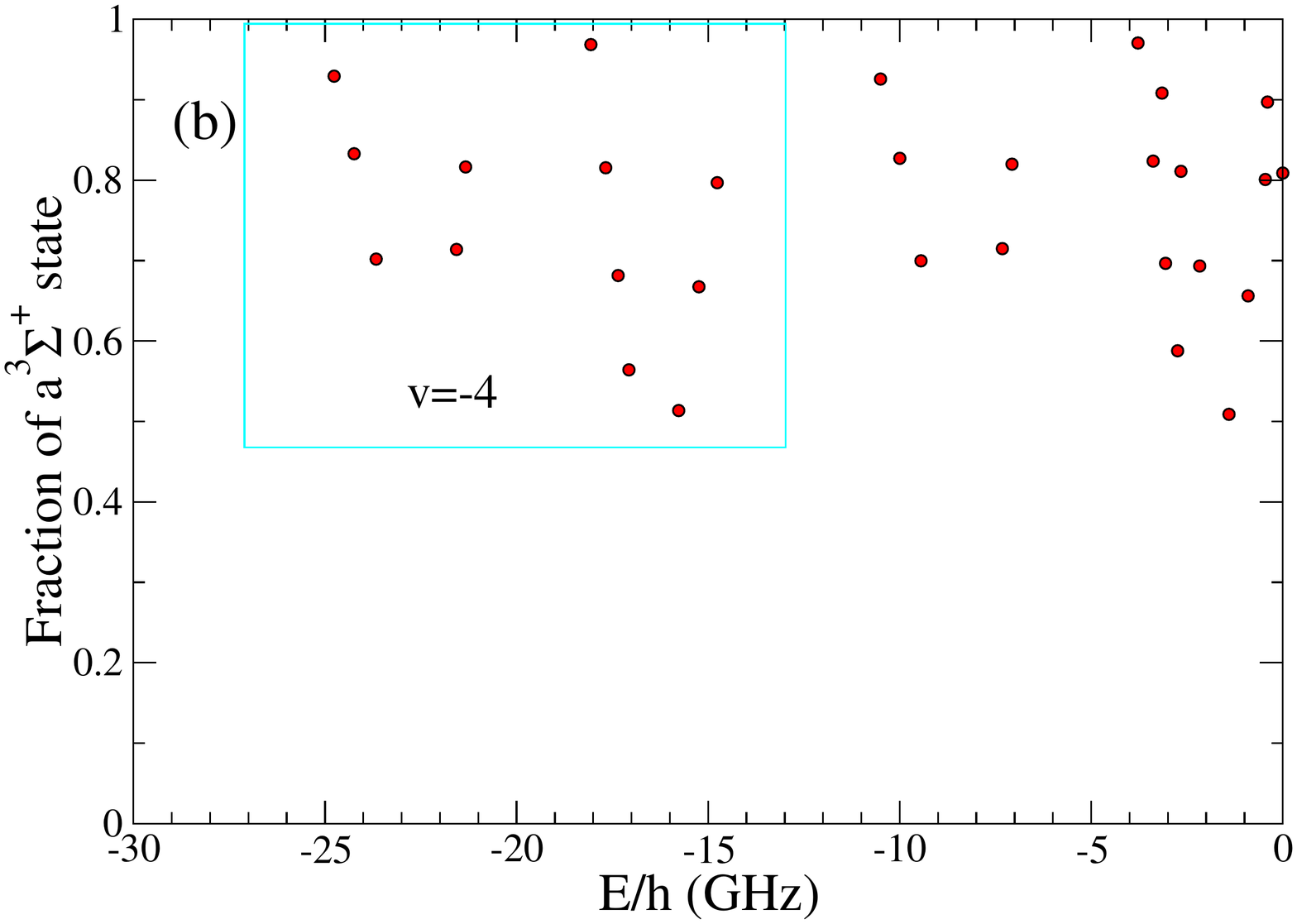}
\caption{Fraction of the a$^3\Sigma^+$ character of the coupled-channel
eigen states of the $^{40}$K$^{87}$Rb molecule as a function of their
energy at a magnetic field of $B$=545.9 G. Bound states with $M_F=-7/2$
and $\ell=0$ are shown. For energy between $-300$GHz and $-15$GHz the
levels can be grouped by vibrational quantum number $v$ of the uncoupled
X$^1\Sigma^+$ and a$^3\Sigma^+$ potentials with $v=-1$ corresponding to
the most weakly bound levels. Panel a displays vibrational levels from
$v=-4$ to $v=-9$ and panel b is blowup of the near threshold region. Zero
energy corresponds to the dissociation energy with both $^{40}$K and
$^{87}$Rb in the energetically lowest hyperfine state.}

\label{mixing}
\end{figure}

Figure \ref{mixing} shows the eigen energies and properties of eigen
functions of  a coupled-channel calculation with $M_F=-7/2$ and $\ell=0$
at a magnetic field of $B$=545.9 G near the dissociation limit. The
figure makes evident that mixing between  the singlet X$^1\Sigma^+$
and triplet a$^3\Sigma^+$ states becomes strong within 150 GHz of
the limit. The vertical axis shows the fraction of the a$^3\Sigma^+$
character in the wave function.  A value close to zero (one) corresponds
to a state primarily described as a X$^1\Sigma^+$ (a$^3\Sigma^+$) level.
Our calculations finds non-degenerate hyperfine and Zeeman structure,
which are grouped in Fig.~\ref{mixing} by a vibrational quantum number.
Within each group there are twelve sublevels, which for weak mixing
corresponds to three singlet and nine triplet sublevels. The vibrational
quantum number is labeled according to the vibrational quantum number
of the uncoupled singlet and triplet potential.  The $v$=$-$1 level
corresponds to the last uncoupled bound level. The grouping is valid in
this case because the two potentials have the same long-range dispersion
potential and sufficiently similar scattering lengths that the spacing
between vibrational levels of the two potentials is nearly the same in
the range of energy shown. For binding energies less than 3 GHz the levels
with different vibrational quantum numbers intermix.  The level relevant
for the experiment of \cite{science08}, which is the initial state of both
Raman transitions in Fig.~\ref{scheme}, is the most weakly bound sublevel
with $M_F=-7/2$ and has a 0.23 MHz binding energy at $B$=545.9 G. Here,
this state has 80 $\%$ a$^3\Sigma^+$ character. The weakly bound levels
are now fully discussed in Ref.~\cite{Julienne}.

\begin{figure}
\includegraphics[scale=0.32]{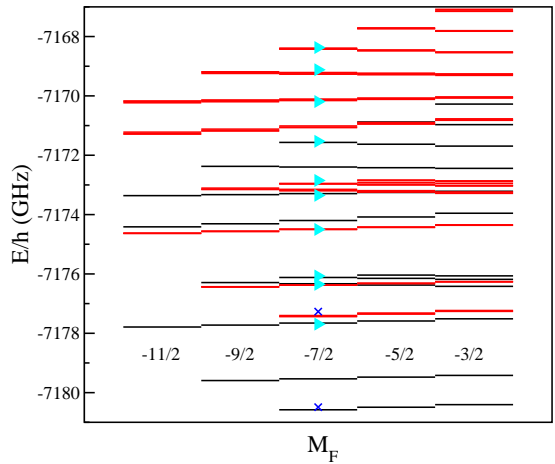}
\caption{The hyperfine and Zeeman structure at $B$=545.9 G of the
$\ell$=0 (black lines) and 2 (red lines) rotational levels of the $v=0$
vibrational state of the a$^3\Sigma^+$ potential of $^{40}$K$^{87}$Rb. The
crosses and triangles indicate the experimentally observed energies
from Ref.~\cite{science08}. The theoretical energies have been shifted
up by +15.0 GHz to coincide with the experimental data. Zero energy
corresponds to the dissociation energy of both $^{40}$K and $^{87}$Rb in
the energetically lowest hyperfine state.  The levels are grouped by the
projection quantum number $M_F$. Each $\ell=2$ hyperfine feature contains
three lines, which on  the scale of the figure are barely resolved. The
splitting is on the order of 0.1 GHz and due to the magnetic spin-spin
dipole interaction, which partially lifts the $m_\ell$ degeneracy of
the projection quantum number of $\vec{\ell}$. }
\label{a_state}
\end{figure}

The final states of the Raman transitions are $v=0$ levels of the
X$^1\Sigma^+$ and a$^3\Sigma^+$ potentials. Figure~\ref{a_state}
shows the rotational hyperfine and Zeeman structure of the $v=0$
level of the a$^3\Sigma^+$ potential at  $B$=545.9 G as calculated
with the coupled-channel method. The black lines are the $\ell$=0 bound states
and the red lines are the $\ell$=2 bound state. Notice though that the
rotational energy splitting is smaller than that due to the hyperfine
and Zeeman interaction. This leads to overlapping  spectral features.
The hyperfine structure of the two partial waves is nearly identical. The
main difference is that each $\ell=2$ hyperfine feature has within it
three lines, which are not resolved in the figure. The splitting between
these lines is less then 0.3 GHz and is due to the
magnetic spin-spin dipole interaction, which partially lifts the $m_\ell$
degeneracy of the projection quantum number of $\vec{\ell}$.

The experiments of Refs.~\cite{nphys08,science08} have located more
than ten sublevels of the $v$=0 vibrational level of the a$^3\Sigma^+$
potential. We have compared the calculated hyperfine structure of
the $v$=0 level of the triplet state, shifted up by +15.0 GHz, with the
experimental energies of Ref.~\cite{science08}, marked by the crosses
and triangles in Fig.~\ref{a_state}. The agreement is good.

\begin{figure}
\includegraphics[scale=0.3]{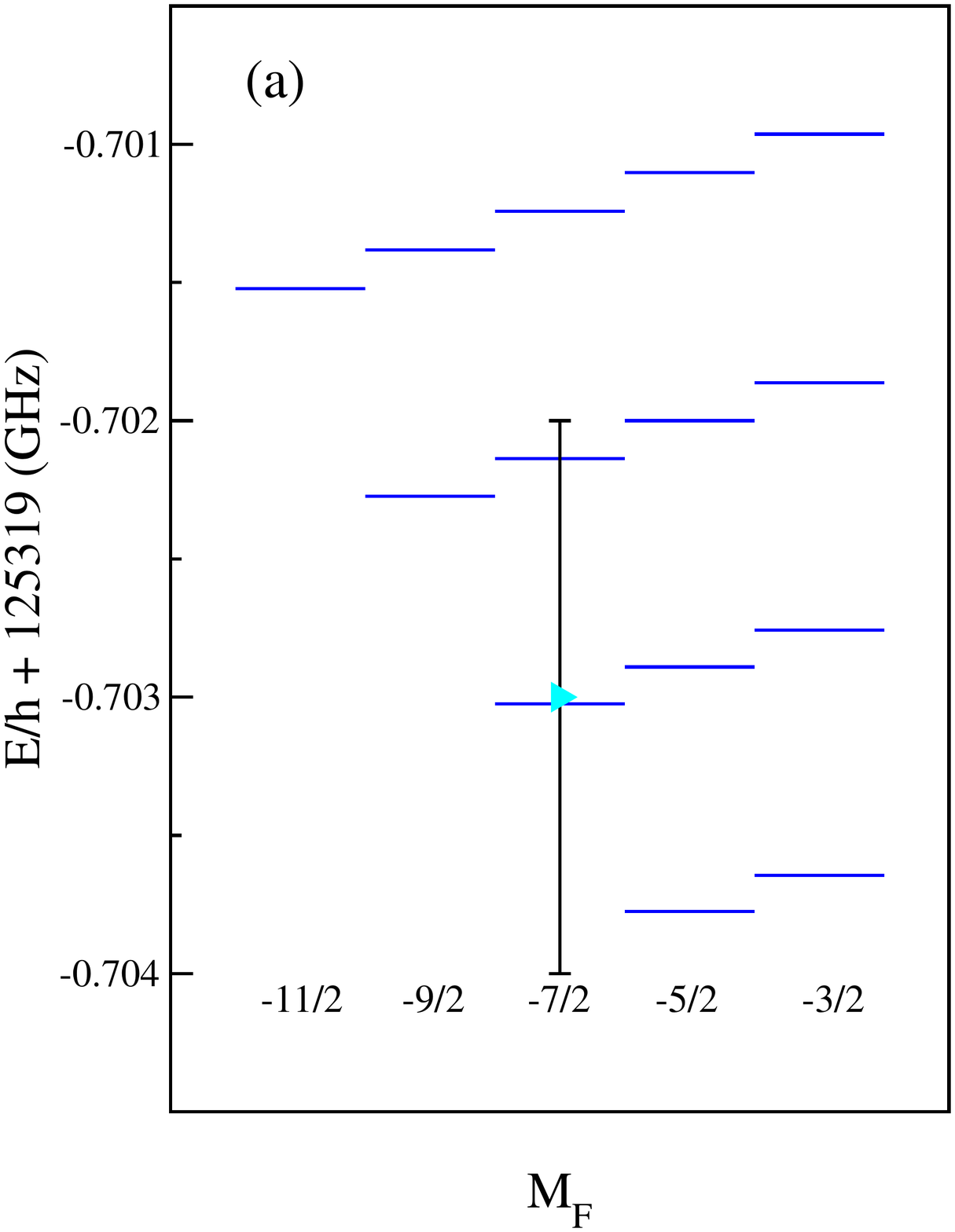}
\includegraphics[scale=0.3]{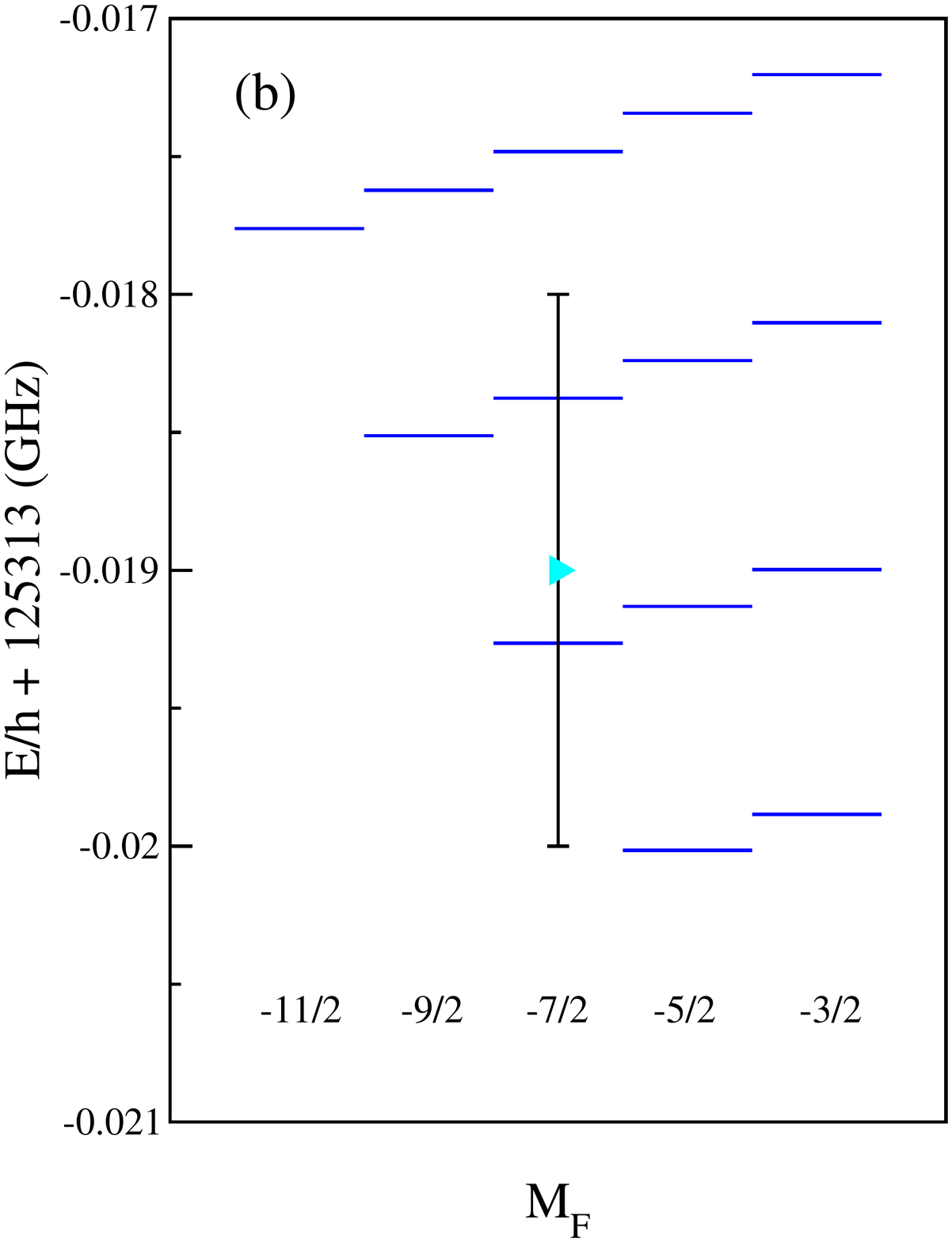}
\caption{The hyperfine and Zeeman structure of the $v$=0 $\ell$=0
level (panel $a$) and $\ell$=2 (panel $b$) of the X$^1\Sigma^+$
state of $^{40}$K$^{87}$Rb at $B=545.9$ G. The triangles with
one-standard-deviation error bar indicate the experimentally observed
energies from Ref.~\cite{science08}. The theoretical energies have been
shifted up by +0.4014 GHz such that the energetically lowest $\ell=0$
$M_F=-7/2$ level coincides with the experimental data. The energy regions
shown in the two panels are not the same.  Zero energy corresponds to the
dissociation energy of both $^{40}$K and $^{87}$Rb in the energetically
lowest hyperfine state. The levels are grouped by the projection quantum
number $M_F$. }
\label{Zeeman_J0J2}
\end{figure}

Figure \ref{Zeeman_J0J2} shows the hyperfine and Zeeman structure
of the $v$=0 vibrational state of the ground X$^1\Sigma^+$ state of
$^{40}$K$^{87}$Rb at $B$=545.9 G shifted up to +0.4014 GHz such that
the energetically lowest $\ell=0$ $M_F=-7/2$ level coincides with the
experimental data indicated by triangles in Fig.~\ref{Zeeman_J0J2}.
The two panel show energy levels for the lowest two even partial waves.
The singlet potential has to first order no hyperfine structure due to the
Fermi contact interaction and electronic Zeeman interaction.  Hence the
structure in Fig.~\ref{Zeeman_J0J2} is predominantly due to the nuclear
Zeeman interaction of both atoms and is on the order of a few MHz at
$B$=545.9 G. In fact, the nuclear Zeeman energy  for the X$^1\Sigma^+$
state is given by
\begin{equation}
E_Z = - (g_{I,K}m_K + g_{I,Rb}m_{Rb}) \mu_N B \,,
\end{equation}
where $m_K$ and $m_{Rb}$ are the projections of the nuclear spin of
$^{40}$K and $^{87}$Rb along the magnetic field direction, respectively.
We then have $M_F=m_K+m_{Rb}$.  The $g_{I,K}$ and $g_{I,Rb}$ are nuclear
g-factors and $\mu_N$ is the nuclear magneton. The g-factor of K is
about a factor of two smaller than that of Rb.  In Fig.~\ref{Zeeman_J0J2}
each line can be labeled by $m_K$ and $m_{Rb}$. The smaller splittings
between lines correspond to levels with different $m_K$ for the same
$m_{Rb}$. The larger gaps between groups of levels correspond to
different $m_{Rb}$.  Each line of Fig.~\ref{Zeeman_J0J2}, panel b,
contains five unresolved components corresponding to the five $\ell$=2
sublevels. Unlike the $\ell$=2 lines of the a$^3\Sigma^+$ state, the
spin-spin dipole interaction here is zero to the first order and the
second order contribution is very small.

For the X$^1\Sigma^+$ potential Ref.~\cite{science08} has observed a
single hyperfine component for the $\ell$=0 and 2 rotational state  of
the $v$=0 vibrational  level. The experiment and theory agree to $\approx$
401.4 MHz in an energy levels. The theoretical energy difference between
$\ell$=0 and $\ell$=2 components of the $M_F=-7/2$ levels is 6.68376 GHz,
which agrees well with the experimental value of 6.6836(5) GHz given
in \cite{science08}.

\section{Multi-channel calculation of the excited states} \label{b}

In this section we model the ro-vibrational motion of the
$^{40}$K$^{87}$Rb molecule in excited electronic potentials, which
are used as intermediates to create vibrationally cold molecules
\cite{science08}.  In particular, we focus on the need to use a
multi-channel description of the vibrational structure that includes
coupling between the electronic potentials. The origin of coupling
can be explained from either the relativistic spin-orbit interaction,
which couples non-relativistic  $^{2S+1}\Lambda^\pm$ Born-Oppenheimer
(BO) potentials \cite{Field} or non-adiabatic mixing of relativistic
$\Omega^\pm$ potentials. Some of the relativistic potentials as function
of internuclear separation $R$ are shown in Fig.~\ref{scheme}.  However,
in this Section we first calculate multi-channel vibrational energies based
on the non-relativistic potentials.  We then discuss the effects of
the multi-channel calculation on the vibrationally-averaged transition
dipole moments. Due to the already complex nature of these calculations,
we have not included the contributions of hyperfine, Zeeman, or coriolis
interactions.

Note that as stated before for a non-relativistic potential the quantum
number $S$, corresponding to the total electron spin $\vec S$, and
$\Lambda$, corresponding the absolute value of the projection of the
total electron orbital angular momentum, is conserved. For a relativistic
calculation only $\Omega$, the absolute value of  the projection of the
summed electronic orbital and spin angular momentum, are conserved. The
$\pm$ superscript, only relevant for $\Lambda=0$ or $\Omega=0$ states,
distinguishing states with opposite reflection symmetries. For a
given $\Omega$ symmetry nonrelativistic potentials that satisfy $-S\le
\Omega-\Lambda \le S$, are coupled by spin-orbit interactions. Strong
mixing occurs when the energy splitting between  $^{2S+1}\Lambda^\pm$
potentials is on the order of the spin-orbit interaction energy.

\begin{figure}
\includegraphics[scale=0.3]{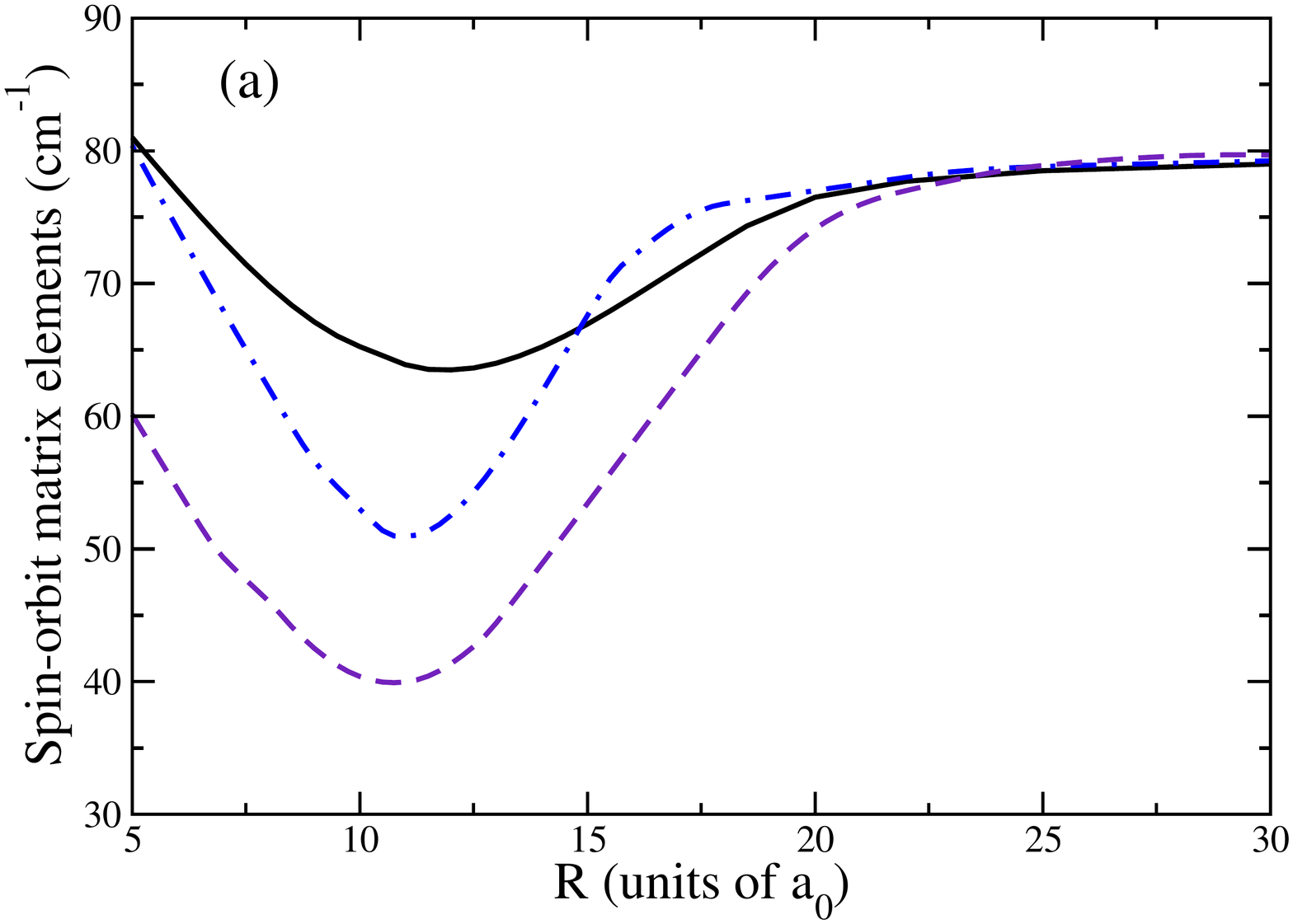}
\includegraphics[scale=0.3]{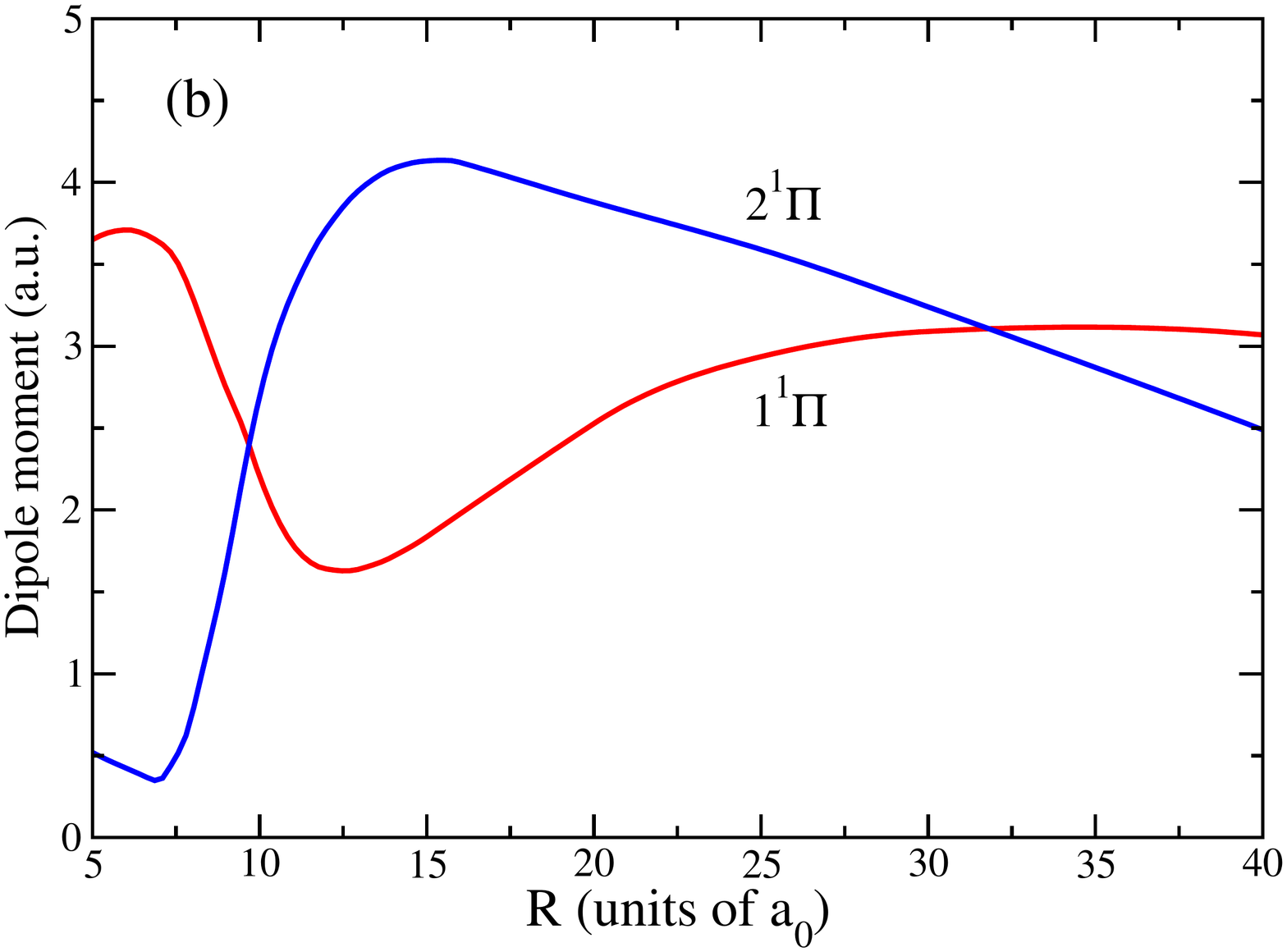}
\caption{Panel a: the $R$-dependent off-diagonal spin-orbit matrix
elements between the $2^3\Sigma^+$ and $1^1\Pi$ potentials (solid line),
the $2^3\Sigma^+$ and $1^3\Pi$ potentials (dashed line) and $1^1\Pi$ and
$1^3\Pi$ potentials (dash-dotted line). For the spin-orbit interaction
between $\Omega=1$ potentials diagonal matrix elements are zero.  Panel b:
the $R$-dependent transition dipole moment between the X$^1\Sigma^+$
and two $^1\Pi$ states. The dipole moments are in units of $ea_0$,
where $e$ is the electron charge.} 
\label{input} \end{figure}

In  our multi-channel calculation we  use the  X$^1\Sigma^+$
non-relativistic ground-state BO potential  from Ref.~\cite{Tiemann}
and the $2^3\Sigma^+$, $1^1\Pi$, and $1^3\Pi$ excited potentials from
Ref.~\cite{Rousseau}. All three excited potentials dissociate to the
K($^2$S)+Rb($^2$P) limit. The previously unknown spin-orbit coupling
matrix elements and electronic dipole moments are obtained from our
MOL-RAS-CI calculations. The panel a in Fig.~\ref{input}  shows the
$R$-dependent spin-orbit matrix elements between the $2^3\Sigma^+$,
$2^1\Pi$, and $1^3\Pi$ states.  For the spin-orbit interaction between
$\Omega$=1 potentials diagonal matrix elements are zero.  As we will be
interested in vibrational levels near the bottom of the 3(1) potential
spin-orbit coupling to $2^1\Pi$ potential can be neglected. At large
$R$ the matrix elements approach $\Delta/3$, where $\Delta$ is the
spin-orbit splitting of the $^2$P state of $^{87}$Rb.  The panel b in
Fig.~\ref{input} shows the $R$-dependent transition dipole moment between
the X$^1\Sigma^+$ and the $1^1\Pi$ and $2^1\Pi$ states. The dipole
moments between the singlet X$^1\Sigma^+$ and the  triplet $2^3\Sigma^+$
and $1^3\Pi$ states are strictly zero.

\begin{figure}
\includegraphics[scale=0.4]{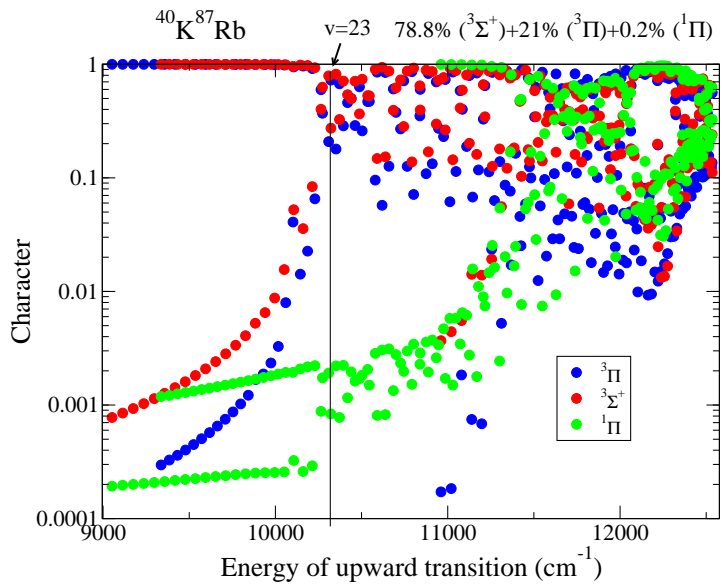}
\caption{The fraction of $^3\Pi$ (blue), $^3\Sigma^+$ (red), and $^1\Pi$
(green) character of the multi-channel $\Omega=1$ eigen states of the
$^{40}$K$^{87}$Rb molecule as a function of energy.  The level indicated
by the label $v=23$ corresponds to the $v$=23 bound state of the 3(1)
potential used in \cite{science08}. Zero energy corresponds to the
dissociation energy of both $^{40}$K and $^{87}$Rb in the energetically
lowest hyperfine state. Hence the energy can be interpreted as the photon
energy needed to make upward transition  shown in the right panel of
Fig.~\ref{scheme}.}
\label{Mchannel} 
\end{figure}

Figure \ref{Mchannel} shows the results of a multi-channel calculation
of the vibrational levels of the $\Omega=1$ excited states. We have
coupled three channels, $2^3\Sigma^+$, $2^1\Pi$, and $1^3\Pi$, and,
in addition, included the rotational potential $\hbar^2J(J+1)/(2\mu
R^2)$ with rotational quantum number $J=\Omega=1$ to each channel. The
vibrational energy on the horizontal axis is relative to the dissociation
energy of ground state $^{40}$K and $^{87}$Rb and thus corresponds
to the photon energy for the upward part of the Raman transition.
Each vibrational level is represented by three circles, of different color
but positioned at the same eigen energy, corresponding to the fraction of
the three-channel wavefunction that is in the $2^3\Sigma^+$, $2^1\Pi$,
and $1^3\Pi$ state, respectively.  The sum of these fractions adds up
to one. A  level with a fraction one in a single channel corresponds to
a vibrational level of an unperturbed $^{2S+1}\Lambda^\pm$ potential.

The energy range in Fig.~\ref{Mchannel} spans from just below the
bottom of the 3(1) potential (See Fig.~\ref{scheme}) to the atomic
K($^2$S)+Rb($^2$P$_{3/2}$) limit. The $v=0$ vibrational level of the
3(1) state can be identified at 9200 cm$^{-1}$. For energies $E>9200$
cm$^{-1}$ levels with a large fraction of $^3\Sigma^+$ character appear.
For $E<9200$ cm$^{-1}$ only vibrational levels of the 2(1) or $^3\Pi$
state exist. Similarly at $E\approx10900$ cm$^{-1}$ eigen states with
a large fraction in the $^1\Pi$ state appear. This corresponds to the
bottom of the 4(1) potential.  Interestingly, for all eigen states with
energy $E<10900$ cm$^{-1}$ the levels have a small  amount, $<0.01$,
of $^1\Pi$ character. The 2(1) and 3(1) states have a small $^1\Pi$ 
admixture due to second-order spin-orbit mixing.

In other energy regions the characterization of levels is less clear. For
example  levels with energy larger than $E\approx 10200$ cm$^{-1}$
have non-negligible contributions from the $2^3\Sigma^+$ and $1^3\Pi$
states. These BO potentials of Ref.~\cite{Rousseau} cross at this energy
and the spin-orbit interaction mixes the two symmetries. The intermediate
vibrational level used in Ref.~\cite{science08} and indicated by $v=23$
in Fig.~\ref{Mchannel} is such a mixed state. From our calculation we
find that it has a 79\% $^3\Sigma^+$, 21\% $^3\Pi$, and 0.2\% $^1\Pi$
character. The closeness of the $v=23$ level to the avoided crossing
and the theoretical uncertainties in its location make the precise
fractions uncertain.  In fact, a few hundred cm$^{-1}$ upward shift could
potentially remove all  $^3\Sigma^+$ and $^3\Pi$ mixing and the $v=23$
level becomes a nearly pure $2^3\Sigma^+$  vibrational level. The $^1\Pi$
character, however, is not expected to change significantly.

\begin{figure}
\includegraphics[scale=0.4]{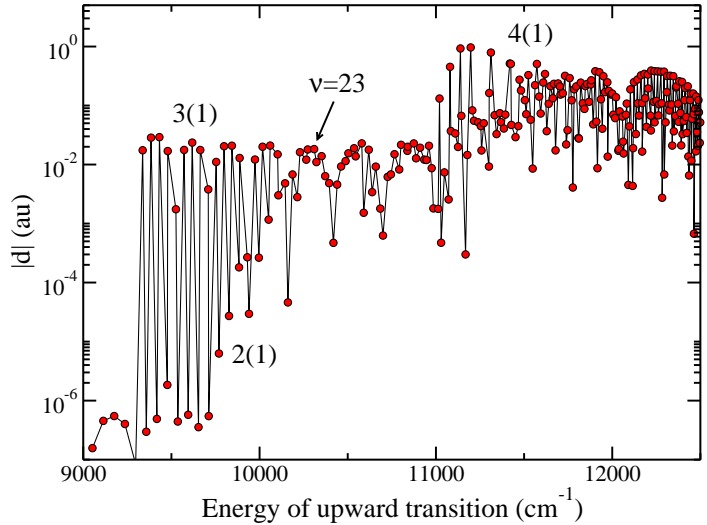}
\caption{Transition dipole moment from the $v$=0, $J$=0 X$^1\Sigma^+$
level to the excited $\Omega$=1 levels of the $^{40}$K$^{87}$Rb molecule
as a function of the excited-state energy. Zero energy corresponds to the
dissociation energy of both $^{40}$K and $^{87}$Rb in the energetically
lowest hyperfine state.  The bound $v$=23 level of the 3(1) potential,
used as intermediate state in \cite{science08}, is marked. The calculated
dipole moment to the $v$=23 level is 0.018 a.u.}
\label{Mdipole}
\end{figure}

Figure~\ref{Mdipole} shows the transition dipole moments between
vibrational levels of the multi-channel $\Omega=1$ calculation
described above and the $v=0$ $J=\ell=0$ ro-vibrational level of
the X$^1\Sigma^+$ potential assuming the electronic dipole moment
shown in Fig.~\ref{input}. This vibrationally-averaged dipole moment
describes the downward part of the Raman transition.  The energy
region is as in Fig.~\ref{Mchannel} and the changing character of the
excited state vibrational levels is reflected in the transition dipole moments.
The vibrational averaged dipole moment from the singlet X$^1\Sigma^+$
state is only nonzero if the multi-channel vibrational levels contains
$^1\Pi$ character. A larger character leads to a larger transition
dipole moment. The start of the vibrational series of the 3(1) and 4(1)
potential are clearly visible in the figure. The intermediate vibrational
level used in Ref.~\cite{science08} is again indicated by $v=23$. From
our calculation we find a dipole moment of 0.018 a.u. for the transition
from this level.

\section{Dynamic polarizability of the ground state vibrational levels}\label{c}

We examine the dynamic polarizability $\alpha$ of the KRb molecule as
a function of laser frequency, $\omega$,  and ro-vibrational quantum
numbers, $v,J$, of the uncoupled ground state a$^3\Sigma^+$ and
X$^1\Sigma^+$ potentials. The real part of the dynamic polarizability,
among other things, determines the depth of the trapping potential seen
by a molecule as
\begin{equation}
            V_0 = -{\rm Re}(\alpha(\hbar\omega,v))\times I\, ,
\end{equation}
where $I$ is the intensity of the laser fields at frequency $\omega$.
The imaginary part of $\alpha$ describes the spontaneous or any other decay
mechanism that leads to loss of molecules from the trap. 
Based on its knowledge, laser frequencies can be selected to minimize decoherence effects from loss of molecules due to spontaneous or laser-induced transitions.

The dynamic polarizibility of a ro-vibrational level of the ground
state is  due to dipole coupling to all other ro-vibronic states of
the ground and excited potentials.  In contrast to the calculations in
Sections~\ref{a} and \ref{b} we do not include multi-channel effects due
to either the hyperfine, Zeeman, or spin-orbit interaction.  Instead,
we base the calculation on our relativistic configuration-interaction
MOL-RAS-CI determination of adiabatic $n(\Omega^\pm)$ potentials and
relativistic transition dipole moments $d(R)$ between the ground-
and excited states.  The relativistic configuration-interaction theory
treats the  spin-orbit interaction non-pertubatively for the electronic
wavefunction.  Consequently, we use for the polarizability
\begin{eqnarray}
    \alpha(\hbar\omega) &=&
\frac{1}{4\pi\epsilon_0}
   \frac{2\pi}{c} 
   \sum_{\Omega'\,v'J'M'}   |\langle \Omega'\, v' J' M' |
                          d(R) \hat{R}\cdot \vec{\epsilon}
                               | \Omega\, vJM \rangle|^2
    \label{eqpolar} \\
    && \quad \times
    \left\{
      \frac{1} { E_{\Omega'v'J'} - i\gamma_{\Omega'v'J'}/2 - ( E_{\Omega vJ}+\hbar\omega )}
    +
      \frac{1} { (E_{\Omega'v'J'} - i\gamma_{\Omega'v'J'}/2 +\hbar\omega) -  E_{\Omega vJ}}
    \right\}
 \nonumber
\end{eqnarray}
where $\hat{R}$ is the orientation of the interatomic axis,
$|\Omega vJM\rangle$ and $|\Omega'v'J'M'\rangle$ are the ro-vibrational
wavefunctions of initial $\Omega$ and final $\Omega'$ states,
respectively. Here, $M$ and $M'$ are the projections of $\vec{J}$ and
$\vec{J}'$ along a laboratory fixed axis.  The vector $\vec{\epsilon}$
is the polarization of the laser, $E_{\Omega vJ}$ is the ro-vibrational
energy in the ground $\Omega$ state and $E_{\Omega'v'J'}$ is the
ro-vibrational energy of the excited $\Omega'$ states.  Contributions from
scattering states or continuum of the excited $\Omega'$ states are also
included.  The widths $\gamma_{\Omega'v'J'}$ describe the spontaneous
decay rate.

In our calculation of the polarizability dipole transitions
to ro-vibrational levels within the a$^3\Sigma^+$ or $X^1\Sigma^+$
potentials as well as to ro-vibrational levels of excited 2(0$^{+/-}$),
3(0$^{+/-}$), 4(0$^{+/-}$), 5(0$^{+/-}$),  2(1), 3(1), 4(1), 5(1) and
6(1) potentials are included.  The 0$^-$ states do not contribute to
the polarizability of the $X^1\Sigma^+$ ro-vibrational levels.

\begin{figure}
\includegraphics[scale=0.4]{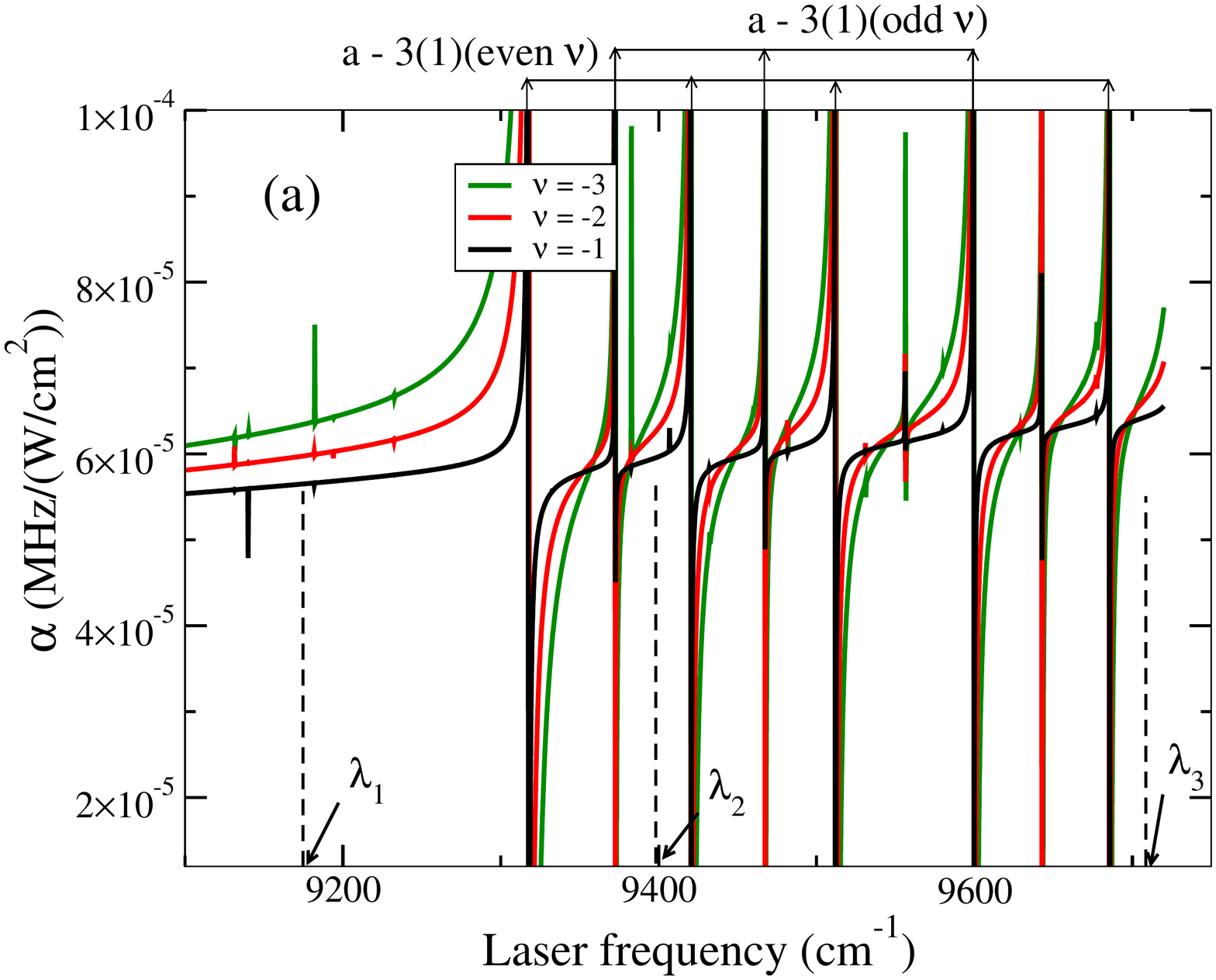}
\includegraphics[scale=0.4]{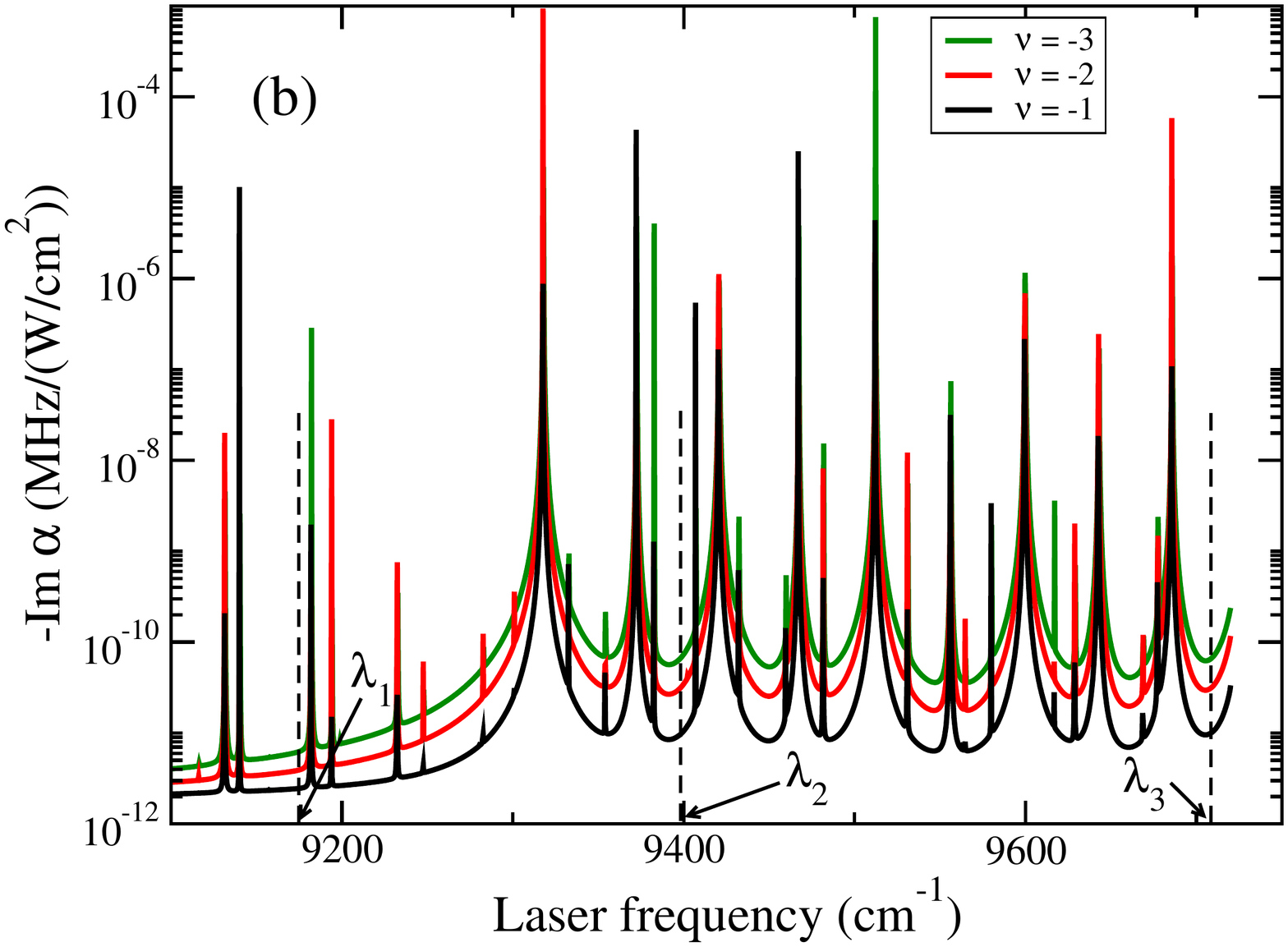}
\caption{
Real (panel a) and minus the imaginary (panel b) part of the dynamic
polarizability $\alpha$ of the $v$=$-$1, $-$2, and $-$3,  $J=0$
ro-vibrational levels of the a$^3\Sigma^+$ ground state of KRb as
a function of laser frequency. The most-widely-used trapping laser
wavelengths, $\lambda_1$ = 1090 nm, $\lambda_2$ = 1064 nm, and $\lambda_3$
= 1030 nm, are indicated.}
\label{polar1_a}
\end{figure}

If a molecule is in a ro-vibrational level of the ground electronic
potential and the frequency of a laser generated optical trap is nearly
resonant to some molecular transition, this will lead to transfer
of population  to the ro-vibrational level of an excited potential,
which then by the spontaneous emission can decay to many ro-vibrational
levels of the ground potential. As result, we lose control over the
molecule in the trap.  To avoid this  we must select  trap frequency
intervals in which resonant excitation is unlikely.  We focus on the
dynamic polarizability of the initial and finals levels relevant to the
Raman transition shown in the right panel of Fig.~\ref{scheme}. Most
experimentally used trapping laser use frequencies that are located in the range
between 9000 cm$^{-1}$ and 9800 cm$^{-1}$. In particular, we examine
the polarizability at laser wavelengths of 1090 nm, 1064 nm, and 1030 nm.

Figure~\ref{polar1_a}  shows the real and imaginary part of the
polarizability of the last three $J=0$ ro-vibrational levels ($v$ =
$-1,-2,-3$) of the a$^3\Sigma^+$ potential as a function of laser
frequency. The polarizability has only been evaluated every 0.05
cm$^{-1}$.  Most of the resonances in Fig.~\ref{polar1_a} are due
to vibrational levels of the 3(1) potential. In the figure we have
assigned the resonances due to transitions from the $v = -1$ level of the
a$^3\Sigma^+$ potential to $even$ or $odd$ vibrational levels of the 3(1)
excited potential. The even $v$ levels have a stronger dipole moment and
thus wider resonance.  Away from the resonances the imaginary part of
the polarizability is six to seven orders of magnitude smaller than the
real part. This indicates that loss due to spontaneous emission from the
excited state is  negligible.  We also indicate three trap wavelengths. As
we can see from Fig.~\ref{polar1_a}, the wavelength $\lambda_1$ = 1090
nm, used in the JILA experiments \cite{nphys08,science08}, is far away
from the resonances and has a very-small decoherence rate. The two other
trap wavelengths, $\lambda_2$ = 1064 nm and $\lambda_3$ = 1030 nm, lie
around stronger resonances and their loss rate is predicted to be ten
to hundred times larger.

\begin{figure}
\includegraphics[scale=0.4]{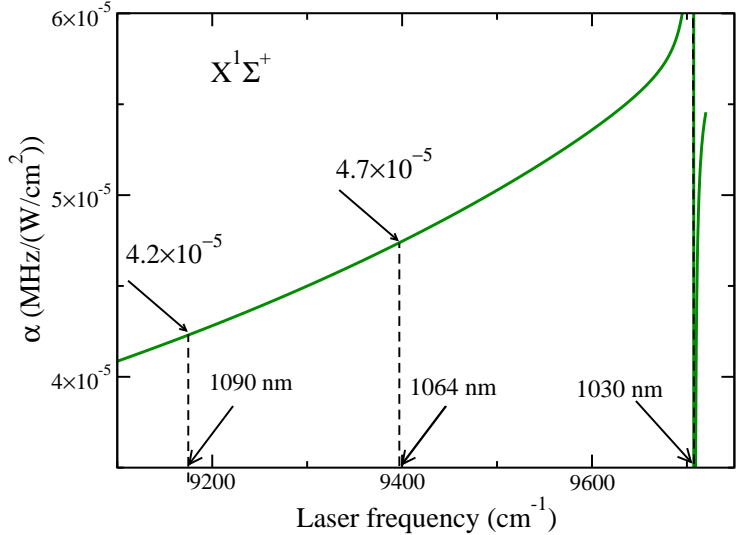}
\caption{
Real part of the dynamic polarizability $\alpha$ of the $v$=0 $J=0$ level
of the X$^1\Sigma^+$ ground state of KRb as a function of laser
frequency within the range of 9100 cm$^{-1}$ to 9710 cm$^{-1}$.}
\label{polar1_X}
\end{figure}

Figure ~\ref{polar1_X}  shows the real part of the polarizability of
the $v$=0 $J=0$ level of the X$^1\Sigma^+$ state of KRb as a function
of laser frequency. Only one resonance is visible. It is due to the
lowest ro-vibrational level of the excited 1$^3\Pi(0^+)$ potential.
This contribution to the polarizability is due to the spin-orbit mixing
with the 2$^1\Sigma(0^+)$ potential.  The benchmark with the shortest
wavelength lies very close to this resonance. However, the precise
location of the resonance is unknown.

In summary, we performed a spin-coupling treatment that describes
the hyperfine and Zeeman structure of the most weakly-bound and the
most deeply-bound vibrational levels of the ground X$^1\Sigma^+$
and a$^3\Sigma^+$ potentials.  The agreement with the observed
experimental structures \cite{science08} is very good. Detailed structural
understanding and assignment of the vibrationally-cold molecules is key
for the coherent control of their interactions.  In addition, we used a
multi-channel description of the excited vibrational levels that includes
$R$-dependent spin-orbit coupling between multiple non-relativistic
$^{2S+1}\Lambda^\pm$ potentials. The spin-orbit coupling constants  were
obtained from an electronic structure calculation.  Finally, we examined
the dynamic polarizability $\alpha$ of vibrationally cold KRb molecules
as a function of laser frequency $\omega$. Based on this knowledge,
laser frequencies can be selected to minimize decoherence from loss of
molecules due to spontaneous or laser-induced transitions.

\section{Acknowledgment}

SK acknowledges support from ARO and AFOSR and PSJ has partial support from ONR.


\begin{references}

\bibitem{nphys08}S. Ospelkaus, A. Peer, K.-K. Ni, J. J. Zirbel,
B. Neyenhuis, S. Kotochigova, P. S. Julienne, J. Ye, and D. S. Jin, Nature Physics {\bf 4}, 622 (2008).

\bibitem{science08}K.-K. Ni, S. Ospelkaus, M. H. G. de Miranda, A. Peer,
B. Neyenhuis, J. J. Zirbel, S. Kotochigova, P. S. Julienne, D. S. Jin, and
J. Ye, Science {\bf 322}, 231 (2008).

\bibitem{Kotoch03}S. Kotochigova,  P. S. Julienne, and E. Tiesinga,
Phys. Rev. A, {\bf 68}, 022501(2003).

\bibitem{Kotoch04}S. Kotochigova, E. Tiesinga, and P. S. Julienne,
Eur. Phys. J. D {\bf 31}, 189 (2004).

\bibitem{Tiemann}A. Pashov, O. Docenko, M. Tamanis, R. Ferber, H. Kn\"{o}ckel, and E. Tiemann,  Phys. Rev. A {\bf 76}, 022511 (2007).

\bibitem{Stwalley}W. C. Stwalley, Eur. Phys. J. D {\bf 31}, 221Ð225 (2004).

\bibitem{DeMille}J. Sage, S. Sainis, T. Bergeman, and D. DeMille, Phys. Rev. Lett. {\bf 94}, 203001
(2005).

\bibitem{Bergeman}T. Bergeman, A. J. Kerman, J. Sage, S. Sainis, and D. DeMille, Eur. Phys. J. D 
{\bf 31}, 179 (2004).

\bibitem{Kasahara}S. Kasahara, C. Fujiwara, N. Okada, and H. Kat\^{o},
J. Chem. Phys. {\bf 111}, 8857 (1999).

\bibitem{Amiot1}C. Amiot, J. Mol. Spect. {\bf 203}, 126 (2000).

\bibitem{Bussery}B. Bussery, Y. Achkar, and M. Aubert-Fre\c{c}on,
Chem. Phys. {\bf 116}, 319 (1987).  

\bibitem{Stoof} H. T. C. Stoof, J. M. V. A. Koelman, and B. J. Verhaar,  Phys. Rev. B {\bf 38}, 4688  (1988).

\bibitem{Tiesinga} E. Wille, F. M. Spiegelhalder, G. Kerner, D. Naik, A. Trenkwalder, G. Hendl, F. Schreck, R. Grimm, T. G. Tiecke, J. T. M. Walraven, S. J. J. M. F. Kokkelmans, E. Tiesinga, and P. S. Julienne,  Phys. Rev. Lett. {\bf 100}, 053201 (2007).

\bibitem{Mies}F. H. Mies, C. J. Williams, P. S. Julienne, and M. Krauss, J. Nat. Inst. Stand. Techn. {\bf 101}, 521 (1996).

\bibitem{Audi} G. Audi, A. H. Wapstra, and C. Thibault, Nuclear Physics A {\bf 729}, 337 (2003).

\bibitem{Arimondo}E. Arimondo, M. Inguscio, and P. Violino, Rev. Mod. Phys. {\bf 49}, 31 (1977).

\bibitem{Stone}N.J. Stone, Oxford University, preprint (2001) found at the website
http://ie.lbl.gov/toi.html; P. Raghavan, At. Data Nucl. Data Tables {\bf 42}, 189 (1989).

\bibitem{Zirbel}J. J. Zirbel, K-K. Ni, S. Ospelkaus, T. L. Nicholson, M. L. Olsen, P. S. Julienne, 
C. E. Wieman, J. Ye, and D. S. Jin, Phys. Rev. A {\bf 78}, 013416 (2008).

\bibitem{Julienne}P. S. Julienne, arXiv:0812:1233.

\bibitem{Field}H. Lefebvre-Brion and R. W. Field, ``{\it The spectra and dynamics of diatomic molecules}" Elsevier Academic Press, 2004.

\bibitem{Rousseau}S. Rousseau, A. R. Allouche, and M. Aubert-Fre\c{c}on,
J. Mol. Spect. {\bf 203}, 235 (2000).
\end{references}
\end{document}